\documentstyle[aps,prb,psfig,epsfig,preprint]{revtex}
\begin{document}
\draft
%
\title{Network patterns and strength of orbital currents in layered cuprates}
\author{M. Eremin$^{1}$ and I. Eremin$^{1,2}$}
\address{$^1$Physics Department, Kazan State University, 420008 Kazan, Russia}
\address{$^2$Institut f\"ur Theoretische Physik, Freie Universit\"at 
Berlin, D-14195 Berlin, Germany}
\date{\today}
\maketitle
\begin{abstract}
In a frame of the $t-J-G$ model we derive the microscopical 
expression for the circulating orbital currents in layered cuprates 
using the anomalous correlation functions. In agreement with 
$\mu$-on spin relaxation ($\mu$SR), 
nuclear quadrupolar resonance (NQR) and inelastic neutron 
scattering(INS) experiments in YBa$_2$Cu$_3$O$_{6+x}$ 
we successfully explain the order of magnitude and the 
monotonous increase of the {\it internal} 
magnetic fields resulting from these currents upon cooling. 
However, the jump in the intensity of the magnetic fields at T$_c$ 
reported recently seems to indicate a non-mean-field feature in 
the coexistence of current and superconducting states and  
the deviation of the extended charge 
density wave vector instability from its
commensurate value {\bf Q}$\approx$($\pi,\pi$) 
in accordance with the reported topology of the Fermi surface. 
\end{abstract}
%
\narrowtext

A possibility for a staggered orbital current phase formation in layered
cuprates has attracted much interest recently 
\cite{1,2,3,4,5,6,7,8,9,10,11}. Most importantly, it was shown that most 
of the observed properties referred to a pseudogap phenomenon 
can be naturally explained in an extended charge density wave (CDW) scenario 
with a complex order parameter phase formation (shortly $s+id$-CDW) 
in underdoped cuprates. The real $s-$wave symmetry component
corresponds to a conventional charge (or spin) density waves whereas
the imaginary part of the 
order parameter has a $d-$ wave symmetry and corresponds to the  
staggered current phase. Different kind of experiments can be interpreted in
favor of the staggered orbital current phase such as an observation of the 
orbital antiferromagnetism in YBa$_2$Cu$_3$O$_{6+y}$ by means of 
inelastic neutron scattering(INS) experiments reported in Refs. \cite{5,6} 
and zero-field muon spin relaxation ($\mu$sR) experiments\cite{8}.
Moreover, recent investigations using nuclear magnetic resonance (NMR) 
technique indicate the presence of the internal fluctuating magnetic 
fields in the superconducting state of layered cuprates \cite{9,10,11}. 
Most importantly, the observed enhancement of the magnetic moment's 
intensity at $T_c$\cite{5,6} seems to indicate an intrinsic and a
non-trivial relation between the superconducting 
and the pseudogap phases. In this connection a microscopical 
analysis of the network patterns 
and the strength of orbital currents becomes very actual.

In general, the possibility of the 
$s+id$-CDW phase formation is related to a divergence of the dynamical 
charge susceptibility at wave vector 
${\bf Q}_i \approx (\pi,\pi)$ in the first Brillouin zone and was 
demonstrated recently for cuprates\cite{18}. Here we derive the analytical 
expression for the current flow and show how its orbital contour can be 
reconstructed for any arbitrary chosen instability wave vector 
${\bf Q}_i$. Most importantly, we calculate the intensity of the resulting 
internal magnetic fields and the corresponding orbital magnetic moments. 
We find that its enhancement at $T_c$ may result from the presence of a   
relatively small $s-$component of the extended CDW. The latter 
agrees well with an observation of the increase of the 
NQR linewidth at Cu(2) site (see Ref. \cite{11}). 
In addition, the non-mean-field character of the
coexistence of superconductivity and $s+id$-CDW phases has to be taken into 
account.

{\it Hamiltonian and general expression for the current flow.}
In our analysis we start from the following $t-J-G$ model Hamiltonian:
\begin{equation}
H=\sum_{ij} t_{ij}\Psi _i^{pd,\sigma }\Psi _j^{\sigma ,pd}+\sum_{i>j}
J_{ij}\left[({\bf S}_i{\bf S}_j)-\frac{n_in_j}4\right]+\sum_{i>j}
G_{ij}\delta _i\delta _{j\text{ }}
\quad,
\label{eq:ham}
\end{equation}
where $\Psi_i^{\alpha,\beta}=\mid i,\alpha ><i,\beta \mid $ are projecting
Hubbard-like operators. Symbol $pd$ corresponds to a Zhang-Rice singlet 
formation with one hole placed on the copper site whereas the second hole 
is distributed on the neighboring oxygen sites \cite{12}. Here $t_{ij}$ is a 
hopping integral, J$_{ij}$ is a superexchange coupling parameter of copper 
spins, and $\sigma =\pm 1/2$. $\delta _i=$ $\Psi_i^{pd,dp}$ is a 
hole doping operator. As in Ref. \cite{14} we also use the 
parameters of a screened Coulomb repulsion of the doped holes at different 
sites, $G_{ij}$. The quasiparticle energy
dispersion and the correlation functions were calculated in a Roth-type of 
a decoupling scheme for the Green's functions\cite{14,13}.

The network patterns and the strength of the orbital currents can be obtained 
using the charge conservation low
\begin{equation}
\frac \partial {\partial t}\int \rho dV=\int \mathbf{j}d\mathbf{S}
\quad.
\end{equation}
The operator of the fluctuating charge per unit cell with number $i$ is
given by
\begin{equation}
e\Psi _i^{pd,pd}=e\delta _i=\widetilde{\delta _i}
\quad,
\label{eq:2}
\end{equation}
that obeys the equation of motion
\begin{equation}
i\hbar \frac \partial {\partial t}\widetilde{\delta _i}=
\left[\widetilde{\delta _i}H\right]
\quad.
\label{eqmo}
\end{equation}
Calculating the commutator with Hamiltonian (\ref{eq:ham}) 
we arrive to the following expression
\begin{equation}
i\hbar \frac \partial {\partial t}\widetilde{\delta _i}=e\sum
t_{ij}\Psi _i^{pd,\sigma }\Psi _j^{\sigma ,pd}-e\sum t_{ji}\Psi
_j^{pd,\sigma }\Psi _i^{\sigma ,pd}
\quad,
\label{eqmof}
\end{equation}
where the right-hand side of this equation is a current operator. 
In order to calculate its thermodynamic value along the link $<ij>$ 
we make the Fourier transform of Eq. (\ref{eqmof}). Then the probability 
of the hopping from $i$ to $j$ site can be written as:
\begin{equation}
<\Psi _i^{pd,\sigma }\Psi _j^{\sigma ,pd}>=\frac 1N \sum_{\bf k,k'} 
<\Psi_{\bf k}^{pd,\sigma }\Psi_{{\bf k}^{\prime }}^{\sigma ,pd}>\exp
(-i{\bf kR}_i+i{\bf k}^{\prime }{\bf R}_j)
\quad,
\label{uprosh1}
\end{equation}
whereas of the reverse process is given by
\begin{equation}
<\Psi _j^{pd,\sigma }\Psi _i^{\sigma ,pd}>=\frac 1N
\sum_{\bf k,k'}<\Psi_{\bf k}^{pd,\sigma }
\Psi _{{\bf k}^{\prime }}^{\sigma ,pd}> \exp(-i{\bf kR}_j+
i{\bf k}^{\prime }{\bf R}_i)
\quad.
\label{uprosh2}
\end{equation}
Since the hopping integral is a real quantity the current flow will be
proportional to the difference of Eqs. (\ref{uprosh1}) and 
(\ref{uprosh2}):
\begin{eqnarray}
\lefteqn{<\Psi _{i}^{pd,\sigma }\Psi _{j}^{\sigma ,pd}-
\Psi _{j}^{pd,\sigma }\Psi_{i}^{\sigma ,pd}> =} \nonumber\\
&& \frac{1}{N}\sum_{\bf k,k'}<\Psi _{\bf k}^{pd,\sigma }\Psi
_{{\bf k}^{\prime }}^{\sigma ,pd}>\{\exp (-i{\bf kR}_{i}+
i{\bf k}^{\prime }{\bf R}_{j})-\exp
(-i{\bf kR}_{j}+i{\bf k}^{\prime }{\bf R}_{i})\}.
\quad.
\label{real}
\end{eqnarray}
At $T<T^{*}$ one have the following non-zero expectation values:
$<\Psi_{\bf k}^{pd,\sigma }\Psi_{\bf k}^{\sigma ,pd}>$, 
$<\Psi_{\bf k+Q}^{pd,\sigma}\Psi_{\bf k}^{\sigma ,pd}>$ and 
$<\Psi_{\bf k}^{pd,\sigma }\Psi_{\bf k+Q}^{\sigma,pd}>$. 
Since the first one does not contribute, we have
\begin{eqnarray}
\lefteqn{<\Psi _{i}^{pd,\sigma }\Psi _{j}^{\sigma ,pd}-
\Psi_{j}^{pd,\sigma}\Psi_{i}^{\sigma ,pd}>=} \nonumber\\ 
&& \frac{1}{N}\sum_{\bf k,k'}<\Psi_{\bf k+Q}^{pd,\sigma }
\Psi _{\bf k}^{\sigma,pd}> \{\exp [-i({\bf k+Q}){\bf R}_{i}+i{\bf kR}_{j}]-
\exp [-i({\bf k+Q}){\bf R}_{j}+i{\bf kR}_{i}]\} 
+ \nonumber\\
&& +\frac{1}{N}\sum_{\bf k}<\Psi_{\bf k}^{pd,\sigma }\Psi_{\bf k+Q}^{\sigma
,pd}>\{\exp [-i{\bf kR}_{i}+i({\bf k+Q}){\bf R}_{j}]-
\exp [-i{\bf kR}_{j}+i({\bf k+Q}){\bf R}_{i}\}.
\label{long}
\end{eqnarray}
In our case the pseudogap order parameter is expected to be the 
complex $(s+id)$-CDW. Therefore, it is useful to separate the correlation
functions into two parts: 
$Re<\Psi _{\bf k+Q}^{pd,\sigma }\Psi _{\bf k}^{\sigma ,pd}>$
and $Im<\Psi _{\bf k+Q}^{pd,\sigma }\Psi _{\bf k}^{\sigma ,pd}>$.
It is straightforward to right further as
\begin{eqnarray}
\lefteqn{<\Psi _i^{pd,\sigma }\Psi _j^{\sigma ,pd}-\Psi _j^{pd,\sigma }\Psi
_i^{\sigma ,pd}>=}\nonumber\\
&& \frac{2i}{N}\sum_{\bf k }Im<\Psi _{\bf k+Q}^{pd,\sigma }\Psi
_{\bf k}^{\sigma ,pd}>\{\cos [{\bf kR}_{j}-({\bf k+Q}){\bf R}_{i}]-
\cos [{\bf kR}_{i}-({\bf k+Q}){\bf R}_{j}]\} 
+\nonumber\\
&& + \frac{2i}{N}\sum_{\bf k}Re<\Psi _{\bf k+Q}^{pd,\sigma }\Psi
_{\bf k}^{\sigma ,pd}>\{\sin [{\bf kR}_{j}-({\bf k+Q}){\bf R}_{i}]-
\sin [{\bf kR}_{i}-({\bf k+Q}){\bf R}_{j}]\}.
\label{eqlong2}
\end{eqnarray}
For the lattice with a mirror plane symmetry perpendicular to the 
$x$ and $y$ axis the integrals over the first 
Brillouin zone containing $\sin {\bf kR}_{ji}$ vanish. 
Thus, in the functions 
$\cos [{\bf kR}_{j}-({\bf k+Q}){\bf R}_{i}]-\cos [-{\bf kR}_{i}+
({\bf k+Q}){\bf R}_{j}],$
and $\sin [{\bf kR}_{j}-({\bf k+Q}){\bf R}_{i}]-\sin [-{\bf kR}_{i}+
({\bf k+Q}){\bf R}_{j}]$
one can leave 
only their parts $\left[ \cos {\bf QR}_{i}-\cos {\bf QR}_{j}\right] 
\cos {\bf kR}_{ji}$
and $\left[ \sin {\bf QR}_{j}-\sin {\bf QR}_{i}\right] \cos {\bf kR}_{ji}$, 
respectively. Then, the contribution to the current flow along the $x$- axis 
due to the nearest hopping can be calculated:
\begin{eqnarray}
\lefteqn{J^{(1)}= \frac{e}{\hbar}t_{1} 
\frac{2}{N}\left[ \cos {\bf QR}_{i}-\cos
{\bf QR}_{j}\right] \sum_{\bf k}Im<\Psi _{\bf k+Q}^{pd,\sigma }\Psi
_{\bf k}^{\sigma ,pd}>\cos {\bf kR}_{ij}+} \nonumber\\
&& + \frac{e}{\hbar}t_1 \frac{2}{N} \left[ \sin {\bf QR}_{j}-\sin {\bf QR}_{i}
\right] \sum_{\bf k}Re<\Psi _{\bf k+Q}^{pd,\sigma }
\Psi _{\bf k}^{\sigma ,pd}>\cos {\bf kR}_{ij}.
\label{Jij}
\end{eqnarray}
Let us discuss at the beginning the simplest case of ${\bf Q}=(\pi ,\pi )$. 
One can immediately see that the second term of Eq. (\ref{Jij}) vanishes. 
Eq. (\ref{Jij}) allows easy to display the network patterns 
for the different symmetries of the order parameter ($s$-, $d$- and so on). 
Most importantly, for the pure d-wave symmetry order parameter 
$Im<\Psi_{\bf k+Q}^{pd,\sigma }\Psi_{\bf k}^{\sigma ,pd}> 
\sim \cos k_{x}-\cos k_{y}$ (see also Ref. \cite{15}) and hence the 
current network pattern is directly mapped on 
the well-known flux-phase state\cite{17}. 

In general case the period of the current pattern is given by 
$a_{x}=2\pi /Q_{x}$ , $a_{y}=2\pi /Q_{y}$. Note, there are also other 
contributions to the network patterns 
due to the next-nearest- ($t_{2}$) and next-next-nearest- 
($t_{3}$) neighbors hopping. These parameters 
are needed for the describing the 
real Fermi surface (see for example Ref. \onlinecite{norman1}). 
The contribution due to $t_2$ is given by
\begin{eqnarray}
\lefteqn{J^{(2)}=\frac{e}{\hbar }t_{2}\sqrt{2}\frac{2}{N}\left[ \cos
{\bf QR}_{i}-\cos {\bf QR}_{j}\right] \sum_{\bf k} 
Im<\Psi_{\bf k+Q}^{pd,\sigma} \Psi_{\bf k}^{\sigma,pd}>
\cos {\bf kR}_{ij}+} \nonumber\\
&& + \frac{e}{\hbar }t_{2}\sqrt{2}\frac{2}{N} 
\left[ \sin {\bf QR}_{j}-\sin {\bf QR}_{i}
\right] \sum_{\bf k}Re<\Psi_{\bf k+Q}^{pd,\sigma }\Psi_{\bf k}^{\sigma ,pd}>
\cos {\bf kR}_{ij}
\quad,
\label{j2}
\end{eqnarray}
and due to $t_3$ as
\begin{eqnarray}
\lefteqn{J^{(3)}=
\frac{e}{\hbar }t_{3}\frac{2}{N}\left[ \cos {\bf QR}_{i}-
\cos {\bf QR}_{j}\right] \sum_{\bf k}Im<\Psi_{\bf k+Q}^{pd,\sigma }
\Psi_{\bf k}^{\sigma ,pd}>\cos {\bf kR}_{ij}+} \nonumber\\ 
&& + \frac{e}{\hbar} t_{3}
\frac{2}{N}\left[ \sin {\bf QR}_{j}-\sin {\bf QR}_{i}\right] 
\sum_{\bf k}Re<\Psi_{\bf k+Q}^{pd,\sigma }\Psi_{\bf k}^{\sigma ,pd}>
\cos {\bf kR}_{ij}
\quad.
\label{j3}
\end{eqnarray}
Note in Eq. (\ref{j2}) indexes $i$ and $j$ refer to the 
next-nearest neighbors whereas in Eq. (\ref{j3}) 
$i$ and $j$ refer to the next-next-nearest neighbors.  

The required correlation function can be calculated straightforwardly in a 
mean-field-approximation and is given by \cite{15}:
\begin{eqnarray}
\lefteqn{<\Psi _{\bf k+Q}^{pd,\sigma }\Psi_{\bf k}^{\sigma ,pd}>= 
\frac{P_{pd}G_k}4\left[ 
\frac{1}{E_{1k}} \tanh \left(\frac{E_{1k}}{2k_BT}\right)+\frac{1}{E_{2k}} 
\tanh\left(\frac{E_{2k}}{2k_BT}\right) \right] +\frac{P_{pd}}{2(E_{1k}-E_{2k})^2} \times} \nonumber\\ 
&& \times  \left[ \frac{G_k}{2} 
(\varepsilon _k-\varepsilon_{k+Q})^2+\Delta_k \Delta_k^{*}(G_k+G_k^{*}) 
\right] \left [\frac 1{E_{1k}}\tanh \left(\frac{E_{1k}}{2k_BT}\right)-
\frac 1{E_{2k}}\tanh \left(\frac{E_{2k}}{2k_BT}\right) \right]
\label{corr}
\end{eqnarray}
where
\begin{eqnarray}
\lefteqn{E_{1k,2k}^{2}=
\frac{(\varepsilon _{k}+\varepsilon _{k+Q})^{2}}{2}+\Delta
_{k}\Delta_{k}^{\ast }+G_{k}G_{k}^{\ast }\pm } \nonumber\\ 
&& \pm \frac{1}{2}\left[ (\varepsilon_{k}^{2}-\varepsilon_{k+Q}^{2})^{2}+
4G_{k}G_{k}^{\ast }(\varepsilon_{k}+\varepsilon_{k+Q})^{2}+\Delta_{k} 
\Delta_{k}^{\ast}(G_{k}+G_{k}^{\ast })^{2} \right]^{1/2}.
\label{eq}
\end{eqnarray}
Here $G_{k}=S(T)+iG_{0}(T)(\cos k_{x}a-\cos k_{y}a)$ is an extended CDW gap 
and $\Delta_{k}=\frac{\Delta_{0}(T)}{2}(\cos k_{x}a-\cos k_{y}a)$ 
is a superconducting $d$-wave gap. It is important to note that the
real part of the correlation function (\ref{corr}) contains the term which is
proportional to the superconducting gap $\Delta_{k}\Delta_{k}^{\ast }$.
Therefore, at T=T$_c$ the value of the correlation function should display a 
'step'  which can be responsible for the corresponding 'jump' of the current 
flow. Keeping in mind this qualitative idea let us now turn to the 
numerical calculations.

For ${\bf Q} = (\pi, \pi)$ the resulting current 
is manly determined by Eq.(\ref{Jij}). Using the equation for the 
$id-CDW$ order parameters in a mean field
approximation (see for details Ref. \cite{jsuper}) it is straightforward 
to prove that
\begin{equation}
\sum_{\bf k} Im<\Psi_{\bf k+Q}^{pd,\sigma} \Psi_{\bf k}^{\sigma,pd}>
\cos k_xa \approx \frac{G_0(T)}{J_1+2G_1}(1+\delta).
\label{vstavka}
\end{equation}
Here $J_{1,}$ and $G_1$ are parameters of the superexchange and the 
screened Coulomb repulsion of the holes at the nearest copper sites taken 
to be 120 meV and 135 meV respectively.  
$\delta$ is number of holes per one unit cell.
Comparing Eq. (\ref{vstavka}) and Eq. (\ref{Jij}) one sees that 
the temperature dependence of the current strength 
$J^{(1)}$ is almost the same as for the order parameter $G_0(T)$. 
The latter was calculated self-consistently in Ref.\cite{15} and  
as was shown the results are sensitive to the details of competition 
between superconducting state (SC) and $id$-CDW state.
Below T$_c$ the superconductivity tries to push out $id$-CDW state and 
as a consequence the order parameter $G_{0}(T)$ goes down at $T<T_{c}.$ 
We shall describe this effect approximately as
\begin{equation}
G_{0}(T)=4k_{B}T_{c}x\sqrt{1-(T/T^{\ast })^{2}}-y\theta (T_{c}-T)k_{B}T_{c} 
\sqrt{1-(T/T_{c})^{2}}
\label{eq:gap}
\end{equation}
where $x$ and $y$ are the parameters of coexistence of the superconducting 
and the $s+id$-CDW states. Here 
$\Delta _{0}(T)=2.4k_{B}T_{c}\sqrt{1-(T/T_{c})^{2}}$, 
$\theta (T_{c}-T)$ is a usual theta function and $x$ is a parameter that
depends on a doping level. According to the 
mean field calculations \cite{15} at $T^{*} \approx 3 T_{c}$ $x$ is equal 
one and near the optimal doping $i.e$
when $T^{*} \approx T_{c}$, $x\approx 1/4$. 

In more general case one expects that {\bf Q}$\ne (\pi,\pi)$. 
Then the total current 
is a sum of $J^{(1)}+J^{(2)}+J^{(3)}$ and the 
relation between the current flow and the gap is not as simple as 
in Eq.(\ref{eq:gap}). 
The deviation of ${\bf Q}$ 
from $(\pi ,\pi )$ is naturally expected as a consequence of the changes in 
the topology of Fermi surface away from half-filling. This is also can be 
seen from our previous analysis of the dynamical charge susceptibility 
which becomes divergent along the contour around $\mathbf{Q}=$ $(\pi, \pi)$ 
(see for details Ref.\cite{18}). Therefore, in Fig.1 we present the 
results of our calculations for the orbital currents at 
$\tilde{\bf Q}=(\frac{11}{12}\pi ,\frac{11}{12}\pi )$ for three 
different regimes of co-existence of superconductivity and $s+id$-CDW phase. 
The gap equation yields a maximum of the critical mean field temperature 
of $id$-component of the pseudo-gap $(T_d^{*})$ (or maximum of entropy) 
at this $\tilde{\bf Q}$ (see Ref. \cite{jsuper}). 
Our numerical calculations show that the resulting temperature behavior 
of the induced magnetic field which is directly proportional to the 
orbital current is indeed slightly differs from the temperature dependence 
of the $s+id$-CDW gap. One sees from Fig. 1 that the calculated curve 
reproduces well the observed behavior of INS intensity\cite{5,6} 
shown in the inset. The values of the hopping integrals were chosen 
(in meV) t$_1 = 100$, t$_2 = 15$ and t$_3 = 12$. They reproduce well 
the topology of the Fermi surface for underdoped cuprates. 

We also would like to note the following results of our calculations. 
At T=T$_c$ the jump in the current strength is 
reproduced only for the $x$ and $y$ values representing the non-mean 
field character of the coexistence phase between 
$s+id$-CDW and superconductivity.
However, such a mean-field reduction of the 
effective gap at $T\leq T_c$ was clearly demonstrated earlier by tunneling
spectroscopy\cite{ekino}. Therefore weather or not a 'jump' 
at $T\leq T_c$ exist becomes an important issue. For example there is no 
anomalies at T$_c$ in $\mu$SR experiments \cite{6}, 
but this 'jump' is clearly visible according to the 
neutron scattering data \cite{5,6}.

Let us also comment on the importance of the small $s$-component of the 
extended CDW. It was shown previously that relatively small 
$s-$component of the extended CDW is required for the explanation of 
monotonic increase of the 
nuclear quadrupolar resonance(NQR)-Cu(2) linewidth at $T\leq T_c$ in 
YBa$_2$Cu$_3$O$_{6+y}$ \cite{11}. 
In this connection it is also logical to switch the $s-$component on
in our discussion. This component is real and as one can see from
Eqs. (\ref{j2})-(\ref{corr}) contribute to the orbital current strength 
if one takes into account the deviation of instability 
vector ${\bf Q}$ from $(\pi,\pi)$.  
The values of the $s- $ component was taken as in paper \cite{11} 
with critical temperature $T_s^{*}=100K$. 
As one can see the deviation effect together with $s$-component of the 
pseudogap can reproduce well the observed a 'jump' at 
$T\leq T_c$ in Refs. \cite{5,6}.

There is an additional argument in favor of the relevance $s$- component
pseudogap with respect to the 'jump' of the effective magnetic moment at 
$T\leq T_c$. It is connected with the dynamical character of the 
charge-current state. In fact, the discussed internal fields 
are not static. All experiments test the mean squared 
field $\sqrt{<h_{eff}^2>}$ that results due to an averaging  
dependent on how fast those fluctuations are. 
If the $s-$component exist the energy of the sliding $(s+id)-CDW$ 
condensate depends on the phase of the order parameter $G_k$ at 
$T\leq T_c$ and hence the sliding motion becomes decelerate. 
Effectively it can be viewed as an increase of the measured 
mean square internal magnetic fields. The situation can become even 
more complicated because, as it was stressed recently the appearance of 
the $s-$component of CDW leads to a damping of the quasiparticle 
regime\cite{9}. In other words one can say that the 
appearance of the $s-$ components of CDW stimulate the nano-scale 
localization phenomenon. The relation of the disorder to the
problem of the local magnetic fields in YBa$_2$Cu$_3$O$_{6+y}$ from
experimental point of view was stressed also very recently\cite{sonier}.

Finally we note that the current network pattern, 
corresponding to the case of 
${\bf Q}=0$ was discussed by Varma \cite{varma}. We do not touch
this case here because according to the neutron scattering data the 
${\bf Q}$ is about $(\pi, \pi)$. Therefore this case seems to be less 
actual. Further experimental studies are required 
in order to verify the orientation of the observed magnetic moments. 
According to Ref.\cite{5} they are aligned along $c$-axis 
(this is in agreement with our discussed $(s+id)$-CDW scenario), 
whereas  Sidis {\it et al.,}\cite{6} reports 
the polarization of the magnetic moments in the copper-oxygen plane.

In summary, the observed monotonously increasing of the magnetic moment 
by neutron scattering\cite{5,6} 
is well reproduced our calculations if one identifies 
the temperature of $T^{*}$ around 300 K as a critical temperature of 
$(s+id)$-CDW phase formation. 
This value is correlated with those $T^{*}$ that were
reported earlier as a pseudogap temperature in these compounds. 
We argue that the reported jump in the intensity at 
$T=T_c$ can be attributed to the presence of the relatively small   
$s-$component of the pseudogap due to deviation of the CDW 
instability vector {\bf Q} from $(\pi,\pi)$. The latter results from the 
observed topology of the Fermi surface and the corresponding behavior of the 
dynamical charge susceptibility in underdoped cuprates\cite{18}.

We are thankful for stimulating discussions with A. Dooglav, A. Rigamonti, 
and D. Manske.
This work was supported by the Swiss National Science Foundation 
(Grant No. 7SUPJ062258) and partially by the Russian Scientific Council 
on Superconductivity (Project No. 98014-1). I.E. was supported by the 
Alexander von Humboldt Foundation.

\begin{figure}[t]
\centerline{\epsfig{clip=,file=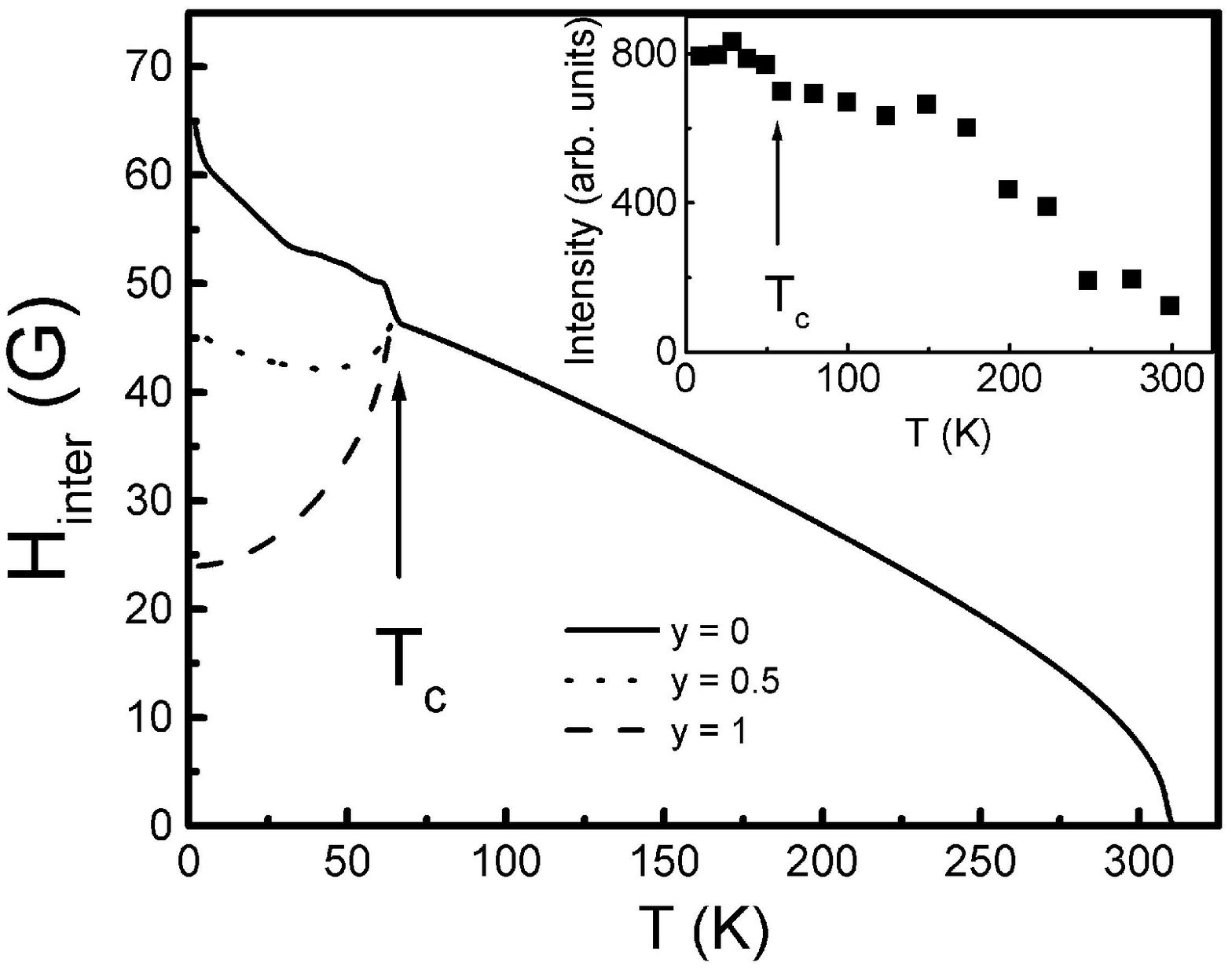,width=13.7cm,angle=0}}
\caption{Calculated intensity of the orbital currents in units of the 
magnetic field\protect\cite{17} 
H$_{int} = \frac{J^{(1)}+J^{(2)}+J^{(3)}}{\hbar c r}$ (where $r \approx 2 \dot{A}$) at $\tilde{\bf Q} = (\frac{11}{12}\pi, \frac{11}{12}\pi)$ 
for three different regimes of the co-existence between extended CDW and 
superconductivity as described in the text. Inset shows the experimental 
data taken from Ref. \protect\onlinecite{5}.}  
\label{fig1}
\end{figure}

\begin{references}
%
\bibitem{1} S. Chakravarty, R.B. Laughlin, D.K. Morr, and Ch. Nayak, 
Phys. Rev. {\bf B63}, 094503 (2001).
%
\bibitem{2} S. Tewari, H.-Y. Kee, Ch. Nayak, and 
S. Chakravarty, Phys. Rev. {\bf B64}, 224516 (2001).
%
\bibitem{3} S. Chakravarty, H.-Y. Kee, and Ch. Nayak, 
Int. J. Mod. Phys. {\bf B 15}, 2901 (2001).
%
\bibitem{4} J.O. Fajerestad and J.B. Marston, cond-mat/0107094 
(unpublished) (2001).
%
\bibitem{5} H.A. Mook, P. Dai, and F. Dogan, Phys Rev. 
{\bf B64}, 012502 (2001).
%
\bibitem{6} Y. Sidis, C. Ulrich, Ph. Bourges, C. Bernhard, 
C.Niedermayer, L.P. Regnault, N.H. Andersen, and B. Keimer, 
Phys Rev. Lett. {\bf 86}, 4100 (2001).
%
\bibitem{7} J.A. Hodges, Y. Sidis, Ph. Bourges, I. Mirebeau, 
M. Hennion, X. Chaud, cond-mat/0107218 (unpublished) (2001).
%
\bibitem{8} J.E. Sonier, J.H. Brewer, R.F. Kiess, R.I. Miller, 
G.D. Morris, C.E. Stronach, J.S. Gardner, S.R. Dunsiger, 
D.A. Bonn, W.N. Hardy, R. Liang, and R.H. Heffner, 
Science {\bf 292}, 1692 (2001).
%
\bibitem{9} M.V. Eremin and A. Rigamonti, Phys. Rev. Lett., to be published
(cond-mat/0103282).
%
\bibitem{10} M.V. Eremin, Yu.A. Sakhratov, A.V. Savinkov {\it et al.}, 
Pis'ma Zh. Eksp. Teor. Fiz. {\bf 73}, 609 (2001) 
$[$JETP Lett. {\bf 73}, 540 (2001)$]$.
%
\bibitem{11} A.V. Dooglav, M.V. Eremin, Yu.A. Sakhratov, and 
A.V. Savinkov, Pis'ma Zh. Eksp. Teor. Fiz. {\bf 74}, 
108 (2001) $[$JETP Lett. {\bf 74}, 103 (2001)$]$.
%
\bibitem{18} M. Eremin, I. Eremin, and S. Varlamov, 
Phys. Rev {\bf B64}, 214512 (2001).
%
\bibitem{12} F.C. Zhang and T.M. Rice, Phys. Rev. {\bf B37}, 3757 (1988).
%
\bibitem{14} M.V. Eremin {\it et al.}, JETP Lett.
{\bf 60}, 125 (1994); J.Phys. Chem. Solids. {\bf 56}, 1713 (1995).
%
\bibitem{13} N.M. Plakida, R. Hayn, and J.L. Richard, Phys. Rev. 
{\bf B51}, 16599 (1995).
%
\bibitem{15} M.V. Eremin and I.A. Larionov, 
Pis'ma Zh. Eksp. Teor. Fiz. {\bf 68} 583 (1998) 
[JETP Lett. {\bf 68}, 611 (1998)].
%
\bibitem{17} T.S. Hsu, J.B. Marston, and I. Affleck, 
Phys. Rev.  {\bf B43}, 2866 (1991).
%
\bibitem{norman1} M.R. Norman, Phys. Rev. {\bf B61}, 14751 (2000).
%
\bibitem{jsuper} I. Eremin and M. Eremin, J. Supercond. {\bf 10}, 
459 (1997); S. Varlamov, M.Eremin, and I. Eremin, 
Pis'ma Zh. Eksp. Teor. Fiz. {\bf 66}, 533 (1997) 
[JETP Lett. {\bf 66}, 569 (1997)].
%
\bibitem{ekino} T. Ekino, Y. Sezaki, and H. Fujii, 
Phys. Rev. {\bf B60}, 6916 (1999). 
%
\bibitem{sonier}  J.E. Sonier, J.H. Brewer, R.F. Kiefl, R.H. Heffner, 
K. Poon, S.L. Stubbs, G.D. Morris, R.I. Miller, W.N. Hardy, 
R. Liang, D.A. Bonn, J.S. Gardner, and N.J. Curro, 
cond-mat/0108479 (unpublished).
%
\bibitem{varma} C. Varma, Phys. Rev. Lett. {\bf 83}, 3538 (1999).
%
%
\end{references}
\end{document}